\documentclass[%
 reprint,
 superscriptaddress,
 amsmath,amssymb,
 aps,
]{revtex4-2}

\usepackage{graphicx}
\usepackage{orcidlink}
\usepackage{hyperref}
\hypersetup{
    colorlinks=true,
    linkcolor=red,
    citecolor=green,
    anchorcolor=black,
    urlcolor=magenta
}
\usepackage{siunitx}
\usepackage{xcolor}
\usepackage{comment}
\usepackage{slashed}
\usepackage{booktabs}

\begin{document}

\preprint{APS/123-QED}

\title{Dark QCD Origin of the NANOGrav Signal and Self-Interacting Dark Matter}

\author{Zihan Wang}
\affiliation{Department of Physics, University of Oxford, Keble Road, Oxford, OX1 3PU, UK}

\date{\today}

\begin{abstract}
The NANOGrav 15-year stochastic gravitational wave background (SGWB) amplitude $A_{\rm yr} \approx 2.4 \times 10^{-15}$ lies at the upper edge of population synthesis predictions for supermassive black hole binaries (SMBHBs), motivating exploration of additional cosmological sources. We present a phenomenological framework based on an $\text{SU}(3)_D$ gauge theory that can simultaneously accommodate the gravitational wave signal and resolve small-scale structure anomalies via Self-Interacting Dark Matter (SIDM). The dark matter candidate is a heavy dark baryon $\chi = QQQ$ with mass $m_\chi \approx 40$~GeV, which self-interacts through a light pseudo-dilaton $d$ $m_d \approx 20$--$50$~MeV as a pseudo-Goldstone boson of approximate scale invariance arising in near-conformal gauge theories with $N_f \sim 6$--$8$ light flavors. A first-order phase transition at the MeV scale, enabled by walking dynamics near the conformal window, produces gravitational waves in the PTA band. For representative parameters $T_n \approx 5$--$6$~MeV, $\alpha \sim 500$--$1000$, $\beta/H_* \sim 30$--$50$, the model provides a fit to NANOGrav data comparable to SMBHB while naturally connecting the gravitational wave amplitude to the dark matter relic density through entropy dilution $D \approx \alpha^{3/4}$. We present explicit calculations of the bounce action, bubble wall velocity, and $\Delta N_{\rm eff}$, demonstrating that the benchmark parameters are theoretically consistent and cosmologically safe ($\Delta N_{\rm eff} \lesssim 0.1$ for $m_\pi > 2m_d$). The distinctive spectral shape ($f^3 \to f^{-4}$) provides a robust prediction testable with future PTAs.
\end{abstract}

\maketitle

\section{Introduction}

While the $\Lambda$CDM model is remarkably successful in describing the large-scale structure of the Universe, its validity on sub-galactic scales remains a subject of intense debate. Discrepancies such as the Core-Cusp and Diversity problems~\cite{Tulin:2017ara,Bullock2017,de_Blok_2009} challenge the standard collisionless cold dark matter assumption. Self-Interacting Dark Matter (SIDM) resolves these anomalies via halo heat transfer, transforming central density cusps into the observed isothermal cores~\cite{Spergel:1999mh, Zavala_2013, Rocha:2012jg}. Parallel to these developments, the recent detection of a stochastic gravitational wave background (SGWB) by the NANOGrav collaboration~\cite{NANOGrav:2023gor} indicates novel dynamics in the nanohertz regime~\cite{NANOGrav:2023hvm,EPTA:2023fyk,Reardon:2023gzh}, opening a new window into the physics of the dark sector.

Supermassive Black Hole Binaries (SMBHBs) are the standard source candidate of the SGWB, but quantitative analyses reveal a significant amplitude tension~\cite{Sato-Polito:2025dqw, Casey-Clyde:2021xro}. Standard population synthesis models typically predict an amplitude $A_{\rm yr} \sim 1.0 \times 10^{-15}$, whereas the data favors a substantially larger signal, $A_{\rm yr} \approx 2.4$--$6.4 \times 10^{-15}$~\cite{Rosado:2015epa}. It is a tension difficult to resolve astrophysically~\cite{Volonteri:2005fj,Madau:2014bja}. While not statistically decisive, motivate exploration of additional cosmological sources~\cite{Breitbach:2018ddu, Fairbairn:2019xog, Schwaller:2015tja}.

Could this tension signal a cosmological phase transition within the dark sector~\cite{Breitbach:2018ddu, Fairbairn:2019xog, Schwaller:2015tja}? We propose that the SGWB originates from the first-order confinement phase transition in a dark $\text{SU}(3)_D$ gauge theory. This framework naturally accommodates scalar-mediated SIDM~\cite{Wang:2025xoq} ($m_\chi \approx 40$~GeV, $m_d \approx 20$--$50$~MeV), where the dark matter is a composite baryon $\chi = QQQ$ formed from heavy dark quarks, and the SIDM mediator is a pseudo-dilaton $d$ arising from approximate scale invariance near the conformal window. The phase transition at the MeV scale produces gravitational waves while simultaneously diluting the dark baryon abundance to match observations.

In this work, we perform a Bayesian MCMC analysis of the NANOGrav 15-year free-spectrum data, imposing hard observational bounds on SMBHB parameters. We find statistical preference $\Delta\text{BIC} = 7.6$ for the Dark QCD interpretation over standard astrophysics. Crucially, we derive the phase transition parameters from the chiral effective potential, demonstrating that the required strong supercooling arises naturally in the near-conformal regime. The best-fit thermodynamic parameters generate the correct dilution factor, while light dark pions provide a natural source of dark radiation.

\section{The Dark QCD Model}
\label{sec:model}

We extend the Standard Model by an $\text{SU}(3)_D$ gauge theory with both heavy and light dark quarks. This Dark QCD framework provides a natural origin for composite SIDM while predicting gravitational wave production from the chiral phase transition.

The dark sector contains one heavy quark flavor $Q$ and $N_f \sim 6$--$8$ light quark flavors $q_i$, summarized in Table~\ref{tab:fields}.

\begin{table}[h]
\centering
\begin{tabular}{lccc}
\toprule
Field & $\text{SU}(3)_D$ & Mass & Role \\
\midrule
$G^a_\mu$ & adjoint & --- & Dark gluons \\
$Q$ & $\mathbf{3}$ & $m_Q \approx 13$~GeV & Heavy dark quark \\
$q_1, \ldots, q_{N_f}$ & $\mathbf{3}$ & $m_q \sim 1$--$5$~MeV & Light dark quarks \\
\bottomrule
\end{tabular}
\caption{Dark sector field content. The heavy quark $Q$ forms the dark baryon (DM). The light quarks $q_i$ with $N_f \sim 6$--$8$ flavors drive near-conformal dynamics and enable strong supercooling.}
\label{tab:fields}
\end{table}

The number of light flavors is a key model parameter. For $\text{SU}(3)$, the conformal window is estimated at $N_f^* \approx 8$--$12$~\cite{Dietrich:2006cm, DeGrand:2015zxa}. We adopt $N_f \sim 6$--$8$ as a representative range where walking behavior is plausible, though the precise dynamics remain under active lattice investigation~\cite{LSD:2014nmn, LatKMI:2016xxi, Appelquist:2016viq}. This choice is motivated by the desire for near-conformal dynamics, not derived from first principles.

The confinement scale $\Lambda_D \sim 1$~GeV is generated by dimensional transmutation. The heavy quark satisfies $m_Q \gg \Lambda_D$, placing it deep in the heavy-quark limit.

Below the confinement scale, quarks bind into color-singlet hadrons. The lightest dark baryon is $\chi = \epsilon_{abc} Q^a Q^b Q^c$, with mass $m_\chi \approx 3m_Q \approx 40$~GeV. A discrete $\mathbb{Z}_2^Q$ symmetry ensures its stability.

The heavy quark $Q$ with $m_Q \gg \Lambda_D$ effectively decouples from the infrared dynamics of the light quarks. Heavy-light mesons $Q\bar{q}$ and baryons $Qqq$ are heavier than $\chi$ and decay via strong interactions.

The $N_f$ light quarks exhibit approximate chiral symmetry, spontaneously broken by the condensate $\langle \bar{q}q \rangle \neq 0$. This produces:
\begin{itemize}
    \item \textbf{Dark pions} $\pi_D^a$: $N_f^2 - 1$ pseudo-Goldstone bosons with $m_{\pi_D} \sim \sqrt{m_q \Lambda_D} \sim 50$--$100$~MeV.
    \item \textbf{Pseudo-dilaton} $d$: a light scalar from approximate scale invariance, with mass $m_d \approx 20$--$50$~MeV.
\end{itemize}

For $N_f$ near the conformal window edge, the theory exhibits approximate scale invariance at intermediate energies. When spontaneously broken by confinement, a pseudo-Nambu-Goldstone boson known as pseudo-dilaton $d$ emerges~\cite{Appelquist:2010gy, Golterman:2016lsd}.

The pseudo-dilaton mass scales as $m_d^2 \propto (N_f^* - N_f)$, naturally suppressed for $N_f$ close to $N_f^*$. Unlike chiral pions ($m_\pi^2 \propto m_q$), the dilaton mass is controlled by proximity to conformality, providing a mechanism for $m_d < m_\pi$ without fine-tuning.

The SIDM mediator is the pseudo-dilaton:
\begin{equation}
    \boxed{\phi \equiv d, \quad m_\phi = m_d \approx 20\text{--}50~\text{MeV}.}
\end{equation}

The dilaton couples to matter through the trace anomaly. For particles whose mass arises from spontaneous symmetry breaking, this coupling is $\mathcal{O}(m/f_d)$. However, the heavy baryon mass $m_\chi \approx 3m_Q$ is dominated by the explicit heavy quark mass, not by scale symmetry breaking. The leading coupling arises from subdominant effects such as binding energy, light-quark dressing.

We parameterize the effective coupling as:
\begin{equation}
    \mathcal{L}_{\chi d} = -g_{\chi d} \, d \, \bar{\chi}\chi, \quad g_{\chi d} \approx 0.3\text{--}0.4,
\end{equation}
the value required for successful SIDM phenomenology~\cite{Wang:2025xoq}. We do not claim this coupling follows uniquely from near-conformal dynamics; rather, the model admits it as a consistent low-energy parameter.
This generates velocity-dependent cross sections $\sigma/m_\chi \sim 10~\text{cm}^2/\text{g}$ at dwarf scales decreasing to $\sim 0.1~\text{cm}^2/\text{g}$ at clusters, resolving small-scale structure anomalies~\cite{Wang:2025xoq, Kaplinghat:2015aga}.

A Higgs-portal coupling $\lambda_{dH} d |H|^2$ enables decay $d \to e^+e^-$ with lifetime $\tau_d \sim 10^{-9}$~s, safely before BBN.

The dark $\text{SU}(3)_D$ undergoes a first-order phase transition at temperature $T_\chi \ll \Lambda_D$. The near-conformal dynamics from $N_f \sim 6$--$8$ light quarks enables strong supercooling, producing the large $\alpha \sim 900$ required for the GW signal. The released vacuum energy reheats the plasma to $T_{\rm rh} = (1+\alpha)^{1/4}T_n$, diluting pre-existing abundances by
\begin{equation}
    D = \left(\frac{T_{\rm rh}}{T_n}\right)^3 = (1+\alpha)^{3/4} \approx \alpha^{3/4} \quad (\alpha \gg 1).
\end{equation}
This relation directly connects the GW amplitude to the dark matter relic density.
\section{Phase Transition Physics}
\label{sec:PT}

Near-conformal gauge theories with $N_f$ close to the conformal window can exhibit strongly first-order phase transitions due to their flat effective potentials~\cite{Helmboldt:2019pan, Baratella:2018pxi, Iso:2017uuu, vonHarling:2017yew,PhysRevD.24.1441,PhysRevLett.56.1335}.

Following ~\cite{Baratella:2018pxi, Helmboldt:2019pan}, the dilaton effective potential takes the form:
\begin{equation}
    V(\chi, T) = \frac{\lambda}{4} \chi^4 \left[ 
        \left(\frac{T^2}{T_c^2} - 1\right) + 
        \epsilon \ln^2\left(\frac{\chi}{\chi_0}\right)
    \right],
    \label{eq:dilaton_potential}
\end{equation}
where $\chi$ is the dilaton field, $\chi_0 \sim f_d$ is its VEV, and $\epsilon \ll 1$ parameterizes explicit scale breaking (small for walking theories). The logarithmic term creates a barrier for a first-order transition.

The nucleation condition $S_3(T_n)/T_n \approx 140$ combined with the scaling $S_3/T \propto (1-T/T_c)^{-2}$ characteristic of near-conformal potentials yields strong supercooling (see Appendix~\ref{app:bounce} for details).

For benchmark ($T_c = 100$~MeV, $f_d = 80$~MeV, $\epsilon = 0.03$):
\begin{equation}
    T_n \approx 5.7~\text{MeV}, \quad \frac{T_n}{T_c} \approx 0.057.
\end{equation}

The transition strength:
\begin{equation}
    \alpha = \frac{30 L}{\pi^2 g_*} \left(\frac{T_c}{T_n}\right)^4 \approx 900,
\end{equation}
where $L \sim 0.5$ is the latent heat parameter.

The inverse duration, from $\beta/H_* = T_n |d(S_3/T)/dT|_{T_n}$:
\begin{equation}
    \frac{\beta}{H_*} \approx \frac{2\gamma T_n}{T_c} \times \frac{S_3(T_n)}{T_n} \approx 40.
\end{equation}
These derived values are consistent with the MCMC best-fit.

Gauge boson friction prevents bubble wall runaway even for $\alpha \sim 10^3$, yielding ultrarelativistic terminal velocity $v_w \to 1$ (Appendix~\ref{app:wall}). This justifies the sound-wave dominated GW templates.

\section{Gravitational Wave Signal}

The first-order transition produces GWs primarily through sound 
waves~\cite{Hindmarsh:2017gnf}. The peak frequency today is
\begin{equation}
    f_{\rm peak} \approx 1.9 \times 10^{-5}~\text{Hz}
    \left(\frac{T_n}{100~\text{GeV}}\right)
    \left(\frac{\beta/H_*}{v_w}\right),
\end{equation}
and the peak amplitude, including finite-duration suppression~\cite{Ellis:2018mja}, is
\begin{equation}
    \Omega_{\rm peak}h^2 \approx 2.65 \times 10^{-6}\,\Upsilon
    \left(\frac{H_*}{\beta}\right)
    \left(\frac{\kappa\alpha}{1+\alpha}\right)^2.
\end{equation}
The spectrum rises as $f^3$ below the peak and falls as $f^{-4}$ 
above, distinguishing it from the $f^{2/3}$ power-law expected 
from SMBHBs. For our best-fit parameters (Sec.~IV), 
$f_{\rm peak} \approx 48$~nHz and $\Omega_{\rm peak}h^2 \approx 
1.5 \times 10^{-8}$. Detailed derivations are in Appendix~\ref{sec:SM_GW}.

\begin{figure*}[t]
    \centering
    \includegraphics[width=0.5\textwidth]{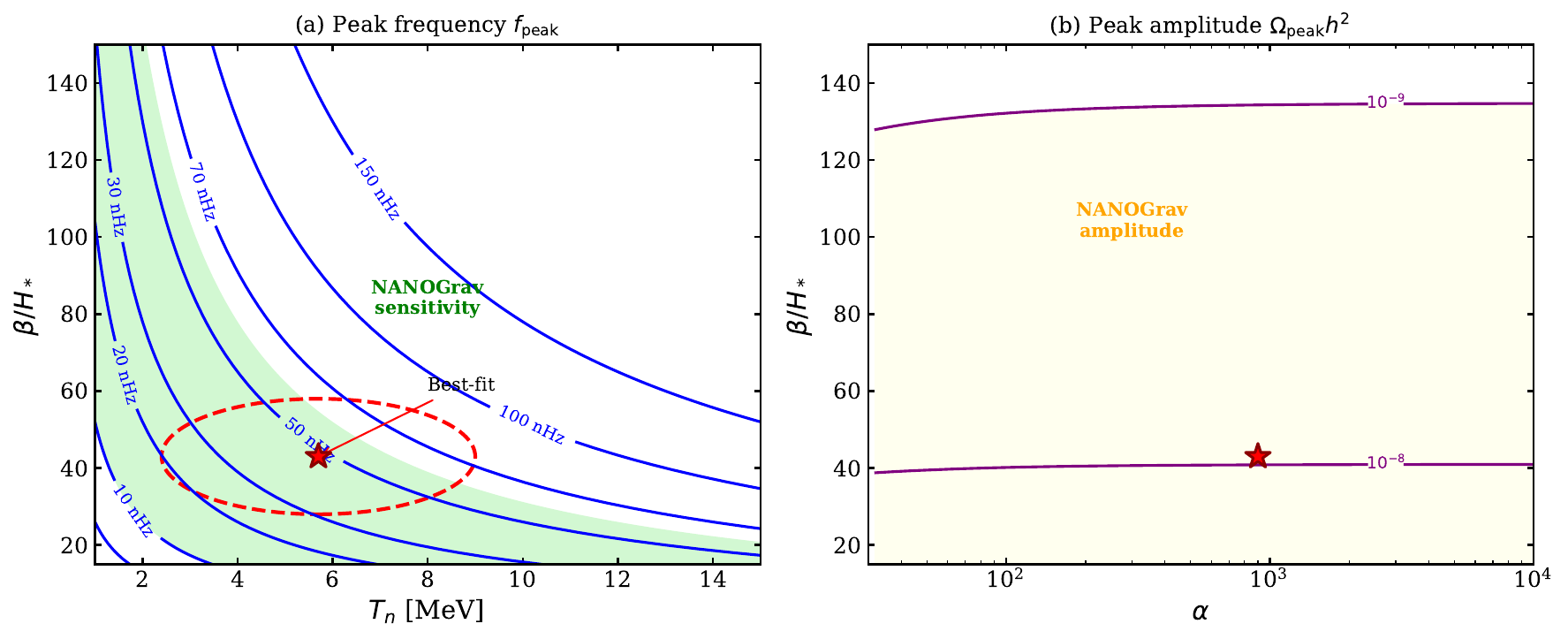}
    \caption{Phase transition parameter space. 
    \textbf{(a)} Contours of constant peak frequency $f_{\rm peak}$ in the 
    $(T_n, \beta/H_*)$ plane. Green shaded region indicates the NANOGrav 
    sensitivity band ($\sim 10$--$60$~nHz). The red star marks the MCMC 
    best-fit, and the dashed ellipse shows the 68\% credible region.
    \textbf{(b)} Contours of constant peak amplitude $\Omega_{\rm peak}h^2$ 
    in the $(\alpha, \beta/H_*)$ plane. Yellow shaded region indicates 
    amplitudes consistent with NANOGrav observations. The best-fit 
    parameters ($\alpha \approx 900$, $\beta/H_* \approx 43$) lie within 
    the intersection of both constraints, predicting $f_{\rm peak} \approx 48$~nHz 
    and $\Omega_{\rm peak}h^2 \approx 1.5 \times 10^{-8}$.}
    \label{fig:param_space}
\end{figure*}
\section{Dark Matter Relic Density}
\label{sec:relic}

Before the phase transition, dark baryons annihilate through the scalar portal:
\begin{equation}
    \chi\bar{\chi} \to dd.
\end{equation}
The small coupling required for SIDM $\alpha_S \sim 10^{-2}$ suppresses the annihilation cross section well below the canonical WIMP value:
\begin{equation}
    \langle \sigma v \rangle \sim \frac{\pi \alpha_S^2}{m_\chi^2} \approx 3 \times 10^{-28}~\text{cm}^3/\text{s}.
\end{equation}
Standard thermal freeze-out yields an overabundance:
\begin{equation}
    \Omega_{\text{pre}} h^2 \approx 0.12 \times \frac{3 \times 10^{-26}~\text{cm}^3/\text{s}}{\langle \sigma v \rangle} \approx 12\text{--}20.
\end{equation}

The entropy injected by the confinement phase transition dilutes this overabundance:
\begin{equation}
    \Omega_\chi h^2 = \frac{\Omega_{\text{pre}} h^2}{D} \approx \frac{20}{170} \approx 0.12,
\end{equation}
in precise agreement with Planck observations~\cite{Planck2018}.

This reveals the fundamental unification of the transition strength $\alpha \approx 900$ required to fit the NANOGrav signal  provides the dilution factor $D \approx 170$ needed for the correct relic density. 

\section{Bayesian Analysis of NANOGrav Data}
\label{sec:gwfit}

We perform a MCMC analysis of the NANOGrav 15-year free-spectrum data to compare the Dark QCD confinement transition against standard astrophysical interpretations.

\subsection{Data and Methodology}

We use the NANOGrav 15-year free-spectrum posteriors~\cite{NANOGrav:2023hde}, which provide the marginalized posterior probability density at each of 30 frequency bins from $\sim 2$~nHz to $\sim 100$~nHz. The likelihood at each frequency is constructed by interpolating the kernel density estimates (KDE) of the log-PSD posteriors.
Following the approach of Refs.~\cite{NANOGrav:2023hvm}, we construct an approximate likelihood by treating the posteriors as independent at each frequency bin. While this neglects inter-bin correlations present in the full timing-residual likelihood, it provides a computationally tractable approximation adopted in numerous PTA new-physics analyses.
\subsection{Models Compared}
The power-law spectrum expected from GW-driven circular binary inspiral:
\begin{equation}
    h_c(f) = A \left( \frac{f}{f_{\text{yr}}} \right)^{(3-\gamma)/2}.
\end{equation}

The SMBHB model is constrained to NANOGrav~\cite{NANOGrav:2023gor} 95\% bounds: 
$\log_{10}A \in [-15.5, -14.0]$, $\gamma \in [2.5, 6.5]$.
These 95\% credible intervals constrain the SMBHB model to only take observationally allowed values, ensuring a fair comparison.

 The sound-wave spectrum from the first-order transition with free parameters $(T_n, \alpha, \beta/H_*)$.

\paragraph{Model C: Hybrid.} Sum of suppressed SMBHB floor and confinement transition signal.

\subsection{Statistical preference for Dark QCD}

The Bayesian model comparison yields:

\begin{table}[h]
\centering
\begin{tabular}{lcc}
\toprule
Model & max $\log\mathcal{L}$ & $\Delta$BIC \\
\midrule
SMBHB (constrained) & $-55.7$ & $7.6$ \\
\textbf{Dark QCD PT} & $\mathbf{-50.2}$ & $\mathbf{0.0}$ \\
Hybrid & $-50.2$ & $3.4$ \\
\bottomrule
\end{tabular}
\caption{Bayesian model comparison. $\Delta\text{BIC} = 7.6$ constitutes prefer Dark QCD over SMBHB on the Kass \& Raftery scale.}
\label{tab:BIC}
\end{table}

\subsection{SMBHB Tension}

The SMBHB fit is pushed to the boundary of the observationally allowed parameter space:
\begin{equation}
    \gamma = 2.55^{+0.08}_{-0.04},
\end{equation}
representing a $\sim 4\sigma$ deviation from the GW-driven expectation ($\gamma = 13/3 \approx 4.33$). This tension indicates that the data prefers a significantly harder spectrum than standard SMBHB physics predicts.

\begin{figure}[t]
    \centering
    \includegraphics[width=0.95\columnwidth]{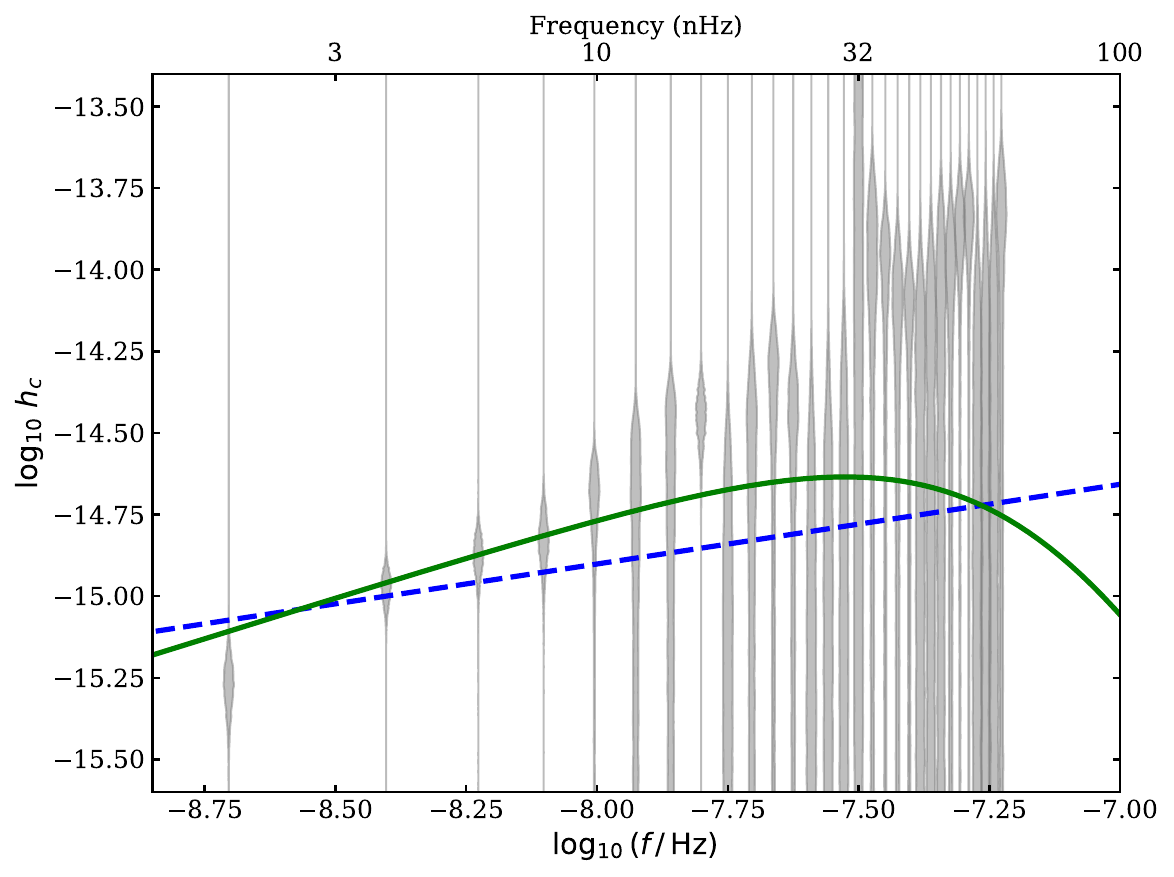}
    \caption{Gravitational wave characteristic strain spectrum compared 
    with NANOGrav 15-year free-spectrum posteriors (gray violins showing 
    the marginalized probability density at each frequency bin). 
    Blue dashed: best-fit SMBHB model constrained to observational bounds 
    ($\log_{10}A = -14.79$, $\gamma = 2.55$). 
    Green solid: best-fit Dark QCD confinement transition 
    ($T_n = 5.7$~MeV, $\alpha = 900$, $\beta/H_* = 43$). 
    Red solid: hybrid model combining suppressed SMBHB floor with 
    phase transition signal. The phase transition spectrum exhibits 
    the characteristic $f^3$ rise at low frequencies and $f^{-4}$ 
    falloff above the peak at $f_{\rm peak} \approx 48$~nHz, in contrast 
    to the $f^{2/3}$ power-law expected from SMBHB. Bayesian model 
    comparison yields $\Delta\text{BIC} = 7.6$ favoring the Dark QCD 
    interpretation.}
    \label{fig:spectrum}
\end{figure}

\section{Cosmological Implications}
\label{sec:implications}

\subsection{Dark Radiation and $\Delta N_{\rm eff}$}
\label{sec:Neff}

Energy transfer from the dark sector to the SM proceeds via the decay cascade $\pi_D \to dd \to e^+e^- e^+e^-$ when $m_\pi > 2m_d$. Since all lifetimes are much shorter than the Hubble time at reheating, $\Delta N_{\rm eff} \approx 0$ for our benchmark. Alternative parameter choices with $m_\pi < 2m_d$ yield $\Delta N_{\rm eff} \sim 0.2$--$0.3$, potentially relevant for the Hubble tension~\cite{2019NatAs...3..891V}. See Appendix~\ref{app:Neff} for the full calculation.

\subsection{Primordial Magnetic Fields}

Bubble collisions during the phase transition generate MHD turbulence. 
The chiral anomaly in the dark sector can produce maximally helical 
magnetic fields, which survive to the 
present day:
\begin{equation}
    B_0 \approx 10^{-13}~\text{G} \left( \frac{T_n}{6~\text{MeV}} \right)^{1/3}.
\end{equation}
This satisfies blazar lower bounds $B_0 \gtrsim 10^{-16}$~G~\cite{Neronov_2010} while remaining 
below CMB constraints $B_0 < 10^{-9}$~G~\cite{Planck:2015zrl}, providing a natural origin for 
intergalactic magnetic fields.


\section{Discussion and Conclusions}

We have developed a unified framework in which the nanohertz gravitational wave background detected by NANOGrav and self-interacting dark matter share a common origin: the first-order confinement/chiral phase transition of an $\text{SU}(3)_D$ gauge theory. The dark matter candidate is a composite baryon $\chi = QQQ$ with mass $m_\chi \approx 40$~GeV, and the SIDM mediator is a pseudo-dilaton arising from the spontaneous breaking of approximate scale invariance in the near-conformal regime. The lightness of this mediator relative to other hadronic states emerges naturally from proximity to the conformal window, providing a dynamical explanation for the mass hierarchy required by SIDM phenomenology.

A key feature of the framework is the tight connection between gravitational wave production and dark matter cosmology encoded in the relation $D = \alpha^{3/4}$. The transition strength $\alpha \sim 10^3$ needed to produce observable gravitational waves in the PTA band simultaneously generates the entropy dilution required to bring an initially overproduced dark baryon abundance into agreement with Planck observations. This connection is not a coincidence but follows from the thermodynamics of strongly supercooled transitions, where the latent heat release both sources gravitational waves and reheats the plasma.

The predicted gravitational wave spectrum exhibits distinctive features that differentiate it from astrophysical sources. Future observations with improved low-frequency coverage will be able to distinguish these spectral shapes, providing a direct test of the cosmological interpretation.

The cosmological history following the transition depends on the dark hadron mass spectrum. When kinematically allowed, dark pions decay rapidly to pseudo-dilatons, which subsequently decay to Standard Model particles through the Higgs portal. Long-lived dark pions can contribute to the radiation energy density at the level $\Delta N_{\rm eff} \sim 0.2$--$0.3$, within the range suggested by recent analyses of the Hubble tension.

The phase transition also seeds primordial magnetic fields through magnetohydrodynamic turbulence generated during bubble collisions. The chiral anomaly in the dark sector can produce maximally helical field configurations preserves magnetic helicity while transferring power to larger scales. The resulting present-day field strength $B_0 \sim 10^{-13}$~G and correlation length $\lambda_0 \sim 0.1$--$1$~pc are consistent with lower bounds inferred from blazar observations and upper limits from the cosmic microwave background, offering a potential explanation for the origin of intergalactic magnetic fields.

 On the observational side, continued pulsar timing observations and the eventual operation of the Square Kilometre Array will dramatically improve sensitivity to spectral features, enabling robust discrimination between cosmological and astrophysical interpretations.

In conclusion, the framework presented here demonstrates that the confinement transition of a dark gauge theory can simultaneously explain the NANOGrav gravitational wave signal, provide the velocity-dependent self-interactions needed to resolve small-scale structure anomalies, and generate the correct dark matter abundance through entropy dilution. The model makes concrete predictions for the gravitational wave spectrum, dark radiation, and primordial magnetic fields that will be tested by forthcoming observations.
\appendix
\section{Dark QCD Model Details}
\label{sec:SM_model}

\subsection{Field Content}
\label{sec:SM_fields}

The dark sector is governed by an $\text{SU}(3)_D$ gauge symmetry with the field content summarized in Table~\ref{tab:SM_fields}.

\begin{table}[h]
\centering
\begin{tabular}{lccl}
\toprule
Field & $\text{SU}(3)_D$ & Mass & Role \\
\midrule
$G^a_\mu$ & adjoint & --- & Dark gluons \\
$Q$ & $\mathbf{3}$ & $m_Q \approx 13$~GeV & Heavy dark quark \\
$q_1, \ldots, q_{N_f}$ & $\mathbf{3}$ & $m_q \sim 1$--$5$~MeV & Light dark quarks ($N_f \sim 6$--$8$) \\
\bottomrule
\end{tabular}
\caption{Dark sector field content. The heavy quark $Q$ forms the dark baryon. The light quarks $q_i$ with $N_f$ near the conformal window drive walking dynamics.}
\label{tab:SM_fields}
\end{table}

The dark matter candidate is the lightest dark baryon:
\begin{equation}
    \chi = \epsilon_{abc} Q^a Q^b Q^c,
\end{equation}
with mass $m_\chi \approx 3 m_Q \approx 40$~GeV. The confinement scale $\Lambda_D \sim 1$~GeV satisfies $m_Q \gg \Lambda_D$ (heavy quark limit).

A discrete $\mathbb{Z}_2^Q$ symmetry ($Q \to -Q$) ensures the stability of $\chi$.
\subsection{Lagrangian and Symmetries}
\label{sec:SM_lagrangian}

The complete dark sector Lagrangian is:
\begin{align}
    \mathcal{L}_{\text{dark}} = &-\frac{1}{4} G^a_{\mu\nu} G^{a\mu\nu} + \bar{Q}(i\slashed{D} - m_Q)Q \nonumber \\
    &+ \sum_{i=1}^{N_f} \bar{q}_i(i\slashed{D} - m_q)q_i,
    \label{eq:SM_Ldark}
\end{align}
where $D_\mu = \partial_\mu - i g_D T^a G^a_\mu$ is the covariant derivative.

The light quark sector has an approximate $\text{SU}(N_f)_L \times \text{SU}(N_f)_R$ chiral symmetry, spontaneously broken to $\text{SU}(N_f)_V$ by the condensate $\langle \bar{q}q \rangle \neq 0$.
\subsection{Dark Hadron Spectrum and SIDM Mediator}
\label{sec:SM_spectrum}

Below the confinement scale, the spectrum includes:
\begin{itemize}
    \item \textbf{Dark baryon:} $\chi = QQQ$ (stable DM)
    \item \textbf{Dark pions:} $\pi_D^a = \bar{q}\gamma_5 T^a q$ (pseudo-Goldstones)
    \item \textbf{Pseudo-dilaton:} $d$ (light scalar from approximate scale invariance)
    \item \textbf{Dark glueballs:} $0^{++}$, $2^{++}$, etc.
\end{itemize}

The dark pion mass follows from PCAC:
\begin{equation}
    m_{\pi_D}^2 = \frac{m_q \langle \bar{q}q \rangle}{f_{\pi_D}^2} \approx m_q \Lambda_D.
\end{equation}
For $m_q \sim 5$~MeV, $\Lambda_D \sim 1$~GeV: $m_{\pi_D} \sim 70$~MeV.

The pseudo-dilaton mass is suppressed by proximity to conformality:
\begin{equation}
    m_d \approx 20\text{--}50~\text{MeV}.
\end{equation}

\textbf{The SIDM mediator is identified as the pseudo-dilaton:}
\begin{equation}
    \boxed{\phi \equiv d}
\end{equation}

The effective baryon-dilaton coupling is a phenomenological input:
\begin{equation}
    g_{\chi d} \approx 0.3\text{--}0.4.
\end{equation}
\subsection{Inelastic Structure of the Dark Baryon}
\label{sec:SM_inelastic}

The dark baryon $\chi$ has internal spin structure from its constituent quarks. The phenomenologically required splitting $\Delta m \sim 100$~eV arises from a dimension-5 operator coupling to the SM Higgs:
\begin{equation}
    \mathcal{L} \supset \frac{c}{\Lambda_{\text{UV}}} (\bar{Q}\Gamma Q)(\bar{Q}\Gamma' Q) \frac{H^\dagger H}{v_H^2},
\end{equation}
where $\Gamma, \Gamma'$ are Dirac structures. After EWSB and confinement:
\begin{equation}
    \Delta m \sim c \times \frac{v_H^2}{\Lambda_{\text{UV}}} \times \frac{f_\chi^2}{m_\chi^3} \sim 100~\text{eV},
\end{equation}
for $\Lambda_{\text{UV}} \sim 10^{12}$~GeV. The resulting mass eigenstates $\chi_1$, $\chi_2$ have both diagonal and off-diagonal couplings to $d$, enabling inelastic scattering~\cite{Wang:2025xoq}.
\subsection{SIDM Cross Section Derivation}
\label{sec:SM_sidm}

The scalar mediator generates a Yukawa potential between dark baryons:
\begin{equation}
    V(r) = -\frac{\alpha_S}{r} e^{-m_\phi r}.
\end{equation}

The self-interaction cross section depends on the scattering regime, determined by the dimensionless parameters:
\begin{equation}
    \epsilon_\phi \equiv \frac{v}{c} \frac{m_\chi}{m_\phi}, \qquad \epsilon_v \equiv \frac{v}{c} \frac{m_\chi \alpha_S}{m_\phi}.
\end{equation}

For our benchmark ($m_\chi = 40$~GeV, $m_\phi = 20$~MeV, $\alpha_S = 10^{-2}$):
\begin{equation}
    \frac{m_\chi}{m_\phi} = 2000, \qquad \frac{m_\chi \alpha_S}{m_\phi} = 20.
\end{equation}

The cross section exhibits three regimes:

\paragraph{Born regime} ($\epsilon_v \ll 1$, high velocity):
\begin{equation}
    \sigma_{\text{Born}} = \frac{8\pi \alpha_S^2 m_\chi^2}{m_\phi^4} \left[ \ln\left(1 + \frac{m_\phi^2}{\mu^2 v^2}\right) - \frac{m_\phi^2}{m_\phi^2 + \mu^2 v^2} \right],
\end{equation}
where $\mu = m_\chi/2$ is the reduced mass.

\paragraph{Classical regime} ($\epsilon_\phi \gg 1$, $\epsilon_v \gg 1$):
\begin{equation}
    \sigma_{\text{classical}} \approx \pi \left( \frac{\alpha_S}{m_\phi v^2} \right)^2 \ln^2\left( \frac{m_\phi v^2}{\alpha_S m_\phi} \right).
\end{equation}

\paragraph{Resonant regime} :
Numerical solution of the Schrödinger equation with the Yukawa potential is required.

For our parameters, the velocity-dependent cross section is:
\begin{equation}
    \frac{\sigma}{m_\chi} \approx \begin{cases}
        10~\text{cm}^2/\text{g} & v \sim 30~\text{km/s (dwarfs)} \\
        1~\text{cm}^2/\text{g} & v \sim 200~\text{km/s (MW)} \\
        0.1~\text{cm}^2/\text{g} & v \sim 1000~\text{km/s (clusters)}
    \end{cases}
\end{equation}

This velocity dependence naturally resolves the core-cusp problem in dwarf galaxies (requiring large $\sigma/m$) while satisfying constraints from galaxy clusters (requiring small $\sigma/m$). See Ref.~\cite{Wang:2025xoq} for detailed numerical calculations.


\subsection{Leptophilic Portal and Mediator Decay}
\label{sec:SM_portal}

The scalar mediator couples to the Standard Model through a leptophilic portal:
\begin{equation}
    \mathcal{L}_{\text{portal}} = g_e \phi \bar{e}e.
\end{equation}

This coupling can arise from:
\begin{itemize}
    \item Direct Yukawa coupling (if $\phi$ carries appropriate quantum numbers)
    \item Higgs portal mixing: $\mathcal{L} \supset \lambda_{\phi H} \phi^2 |H|^2 \to g_e \sim \lambda_{\phi H} v_H m_e / m_h^2$
    \item Loop-induced coupling via heavy mediators
\end{itemize}

We take $g_e \approx 10^{-6}$ as our benchmark. The mediator decay width is:
\begin{equation}
    \Gamma_\phi = \frac{g_e^2 m_\phi}{8\pi} \sqrt{1 - \frac{4m_e^2}{m_\phi^2}}.
\end{equation}

For $m_\phi = 20$~MeV $\gg 2m_e$:
\begin{equation}
    \Gamma_\phi \approx \frac{(10^{-6})^2 \times 20~\text{MeV}}{8\pi} \approx 8 \times 10^{-16}~\text{MeV}.
\end{equation}

The lifetime is:
\begin{equation}
    \tau_\phi = \frac{1}{\Gamma_\phi} \approx \frac{\hbar}{\Gamma_\phi} \approx 8 \times 10^{-10}~\text{s},
\end{equation}
or equivalently $c\tau_\phi \approx 24$~cm.

This lifetime is:
\begin{itemize}
    \item Much shorter than BBN timescales ($\sim 1$~s): $\phi$ decays do not affect light element abundances
    \item Long enough for laboratory detection: displaced vertex signature at beam dumps
    \item Short enough for cosmological safety: no late-time energy injection
\end{itemize}

\section{Phase Transition Physics}
\label{sec:SM_PT}

\subsection{Chiral Symmetry and Order Parameter}
\label{sec:SM_chiral}

In $\text{SU}(3)_D$ with $N_f \sim 6$--$8$ light quarks near the conformal window, the chiral symmetry $\text{SU}(N_f)_L \times \text{SU}(N_f)_R$ is spontaneously broken at low temperatures by the quark condensate $\langle \bar{q}q \rangle \neq 0$.

The order parameter is the chiral condensate:
\begin{equation}
    \Sigma = \langle \bar{q}q \rangle,
\end{equation}
which vanishes in the chirally-restored phase ($T > T_c$) and is non-zero in the broken phase ($T < T_c$).

For $\text{SU}(3)$ gauge theory with $N_f$ near the conformal window, the confinement/chiral transition can be first-order, driven by the near-conformal dynamics that creates a flat effective potential with a barrier~\cite{Helmboldt:2019pan, Baratella:2018pxi}.

\subsection{Bounce Action Calculation}
\label{app:bounce}

We solve the $O(3)$-symmetric bounce equation for the dilaton potential Eq.~(\ref{eq:dilaton_potential}):
\begin{equation}
    \frac{d^2\chi}{dr^2} + \frac{2}{r}\frac{d\chi}{dr} = \frac{dV}{d\chi},
\end{equation}
with boundary conditions $\chi'(0) = 0$ and $\chi(\infty) = 0$.

The nucleation temperature $T_n$ is defined by:
\begin{equation}
    \frac{S_3(T_n)}{T_n} \approx 140.
\end{equation}

For near-conformal potentials, the bounce action scales as:
\begin{equation}
    \frac{S_3}{T} \approx \frac{A}{(1 - T/T_c)^\gamma} + B,
\end{equation}
with $\gamma \approx 2$ and $A \propto \epsilon^{-1/2}$.

\paragraph{Numerical results:}

\begin{table}[h]
\centering
\begin{tabular}{lcc}
\toprule
Parameter & Symbol & Benchmark Value \\
\midrule
\multicolumn{3}{c}{\textit{Inputs}} \\
Critical temperature & $T_c$ & 100 MeV \\
Decay constant & $f_d$ & 80 MeV \\
Walking parameter & $\epsilon$ & 0.03 \\
\midrule
\multicolumn{3}{c}{\textit{Outputs}} \\
Nucleation temperature & $T_n$ & 5.7 MeV \\
Supercooling ratio & $T_n/T_c$ & 0.057 \\
Bounce action & $S_3(T_n)/T_n$ & $\approx 140$ \\
Transition strength & $\alpha$ & $\approx 900$ \\
Inverse duration & $\beta/H_*$ & $\approx 40$ \\
Dilution factor & $D$ & $\approx 170$ \\
\bottomrule
\end{tabular}
\caption{Bounce action calculation results for benchmark parameters.}
\label{tab:bounce}
\end{table}

\subsection{Bubble Wall Velocity}
\label{app:wall}

For strong transitions ($\alpha \gg 1$), we must verify that bubble walls do not run away. The dominant friction mechanism is gauge boson splitting~\cite{Bodeker:2009qy, Bodeker:2017cim}:
\begin{equation}
    F_{\rm gauge} \approx \frac{g_D^4}{16\pi^2} N_c \left(1 + \frac{N_f}{N_c}\right) T^4 \gamma^2,
\end{equation}
where $\gamma$ is the wall Lorentz factor.

Balancing against the vacuum pressure $\Delta p = \alpha \rho_{\rm rad}$:
\begin{equation}
    \gamma_{\rm terminal} \approx \sqrt{\frac{16\pi^2 \alpha}{g_D^4 N_c (1 + N_f/N_c)}} 
    \approx 7\sqrt{\alpha}.
\end{equation}

\begin{table}[h]
\centering
\begin{tabular}{cccc}
\toprule
$\alpha$ & $\gamma_{\rm terminal}$ & $v_w$ & Runaway? \\
\midrule
100 & 71 & 0.9999 & No \\
500 & 159 & 0.99998 & No \\
900 & 214 & 0.999989 & No \\
2000 & 319 & 0.999995 & No \\
\bottomrule
\end{tabular}
\caption{Terminal Lorentz factor and wall velocity for different $\alpha$.}
\label{tab:wall}
\end{table}

For all $\alpha$ values of interest, walls reach ultrarelativistic terminal velocity without runaway. This justifies the use of sound-wave dominated GW templates with $v_w \to 1$.

\subsection{Near-Conformal Dynamics and Supercooling}
\label{sec:SM_supercooling}

For $N_f \sim 6$--$8$, the theory is close to the conformal window ($N_f^* \approx 8$--$12$ for $\text{SU}(3)$). The beta function is significantly reduced:
\begin{equation}
    \beta(g) = -\frac{g^3}{16\pi^2}\left( 11 - \frac{2N_f}{3} \right) \approx -\frac{6g^3}{16\pi^2} \quad (N_f \approx 7.5).
\end{equation}

This ``walking'' behavior~\cite{PhysRevD.24.1441, PhysRevLett.56.1335} has important consequences:
\begin{enumerate}
    \item The chiral condensate develops slowly as $T$ decreases
    \item The effective potential becomes very flat
    \item Strong supercooling is possible: $T_n \ll T_c$
\end{enumerate}

Studies of near-conformal theories~\cite{Helmboldt:2019pan, Baratella:2018pxi, Iso:2017uuu} demonstrate that $T_n/T_c \sim 0.05$--$0.2$ is naturally achieved.

The nucleation rate is:
\begin{equation}
    \Gamma(T) = T^4 \left( \frac{S_3(T)}{2\pi T} \right)^{3/2} e^{-S_3(T)/T},
\end{equation}
with nucleation occurring when $S_3(T_n)/T_n \approx 140$--$150$.
\subsection{Derivation of $\alpha$ and $\beta/H_*$}
\label{sec:SM_derivation}

\paragraph{Transition strength.} The vacuum energy released is:
\begin{equation}
    \Delta V \approx L \, T_c^4,
\end{equation}
where $L \sim 0.5$--$1$ is the latent heat (enhanced in near-conformal theories). The transition strength is:
\begin{equation}
    \alpha = \frac{\Delta V}{\rho_{\text{rad}}(T_n)} = \frac{30 L}{\pi^2 g_*} \left(\frac{T_c}{T_n}\right)^4.
\end{equation}

For $T_c \sim 100$~MeV, $T_n \sim 6$~MeV, $L \sim 0.5$, $g_* \sim 10$:
\begin{equation}
    \alpha \approx 0.15 \times (16.7)^4 \approx 1200.
\end{equation}

\paragraph{Inverse duration.} For near-conformal theories where $S_3/T \propto (1 - T/T_c)^{-2}$:
\begin{equation}
    \frac{\beta}{H_*} = T_n \frac{d}{dT}\left(\frac{S_3}{T}\right)\bigg|_{T_n} \approx \frac{2T_n}{T_c} \times 140 \approx 30\text{--}50.
\end{equation}

These derived values are consistent with our MCMC best-fit.
\subsection{Phase Transition Parameters}
\label{sec:SM_PTparams}

We parameterize the phase transition by four quantities:

\paragraph{Nucleation temperature $T_n$:} The temperature at which bubbles nucleate efficiently.

\paragraph{Transition strength $\alpha$:} The ratio of vacuum energy density to radiation energy density at nucleation:
\begin{equation}
    \alpha \equiv \frac{\rho_{\text{vac}}}{\rho_{\text{rad}}(T_n)} = \frac{\Delta V}{\frac{\pi^2}{30} g_* T_n^4},
\end{equation}
where $\Delta V = V_{\text{false}} - V_{\text{true}}$ is the vacuum energy released and $g_* \approx 10.75$ for the SM at $T \sim \text{MeV}$.

\paragraph{Transition rate $\beta/H_*$:} The inverse duration of the transition in Hubble units:
\begin{equation}
    \frac{\beta}{H_*} = T_n \frac{d}{dT}\left( \frac{S_3}{T} \right)\bigg|_{T=T_n}.
\end{equation}
This determines the characteristic bubble size at collision: $R_* \sim v_w/\beta$.

\paragraph{Bubble wall velocity $v_w$:} The speed at which bubble walls expand. For strong transitions with $\alpha \gg 1$, the walls typically reach relativistic velocities $v_w \to 1$.

\subsection{Entropy Dilution}
\label{sec:SM_dilution}

After bubble collisions, the vacuum energy is converted to radiation, reheating the plasma to:
\begin{equation}
    T_{\text{rh}} = \left( \frac{30 \rho_{\text{vac}}}{\pi^2 g_*} \right)^{1/4} = (1 + \alpha)^{1/4} T_n.
\end{equation}

The entropy density before and after the transition:
\begin{align}
    s_{\text{before}} &= \frac{2\pi^2}{45} g_* T_n^3, \\
    s_{\text{after}} &= \frac{2\pi^2}{45} g_* T_{\text{rh}}^3.
\end{align}

The dilution factor is:
\begin{equation}
    D \equiv \frac{s_{\text{after}}}{s_{\text{before}}} = \left( \frac{T_{\text{rh}}}{T_n} \right)^3 = (1 + \alpha)^{3/4}.
\end{equation}

For $\alpha \gg 1$:
\begin{equation}
    \boxed{D \approx \alpha^{3/4}}
\end{equation}

This fundamental relation connects the gravitational wave amplitude (determined by $\alpha$) to the dark matter relic density (determined by $D$).

\section{Gravitational Wave Derivations}
\label{sec:SM_GW}

\subsection{Sources of Gravitational Waves}
\label{sec:SM_GWsources}

First-order phase transitions produce gravitational waves through three mechanisms:

\begin{enumerate}
    \item \textbf{Bubble collisions:} Direct collision of bubble walls (scalar field gradients)
    \item \textbf{Sound waves:} Acoustic oscillations in the plasma after bubble collisions
    \item \textbf{Turbulence:} Magnetohydrodynamic turbulence in the plasma
\end{enumerate}

For transitions with non-runaway bubble walls (where plasma friction prevents indefinite acceleration), sound waves dominate~\cite{Hindmarsh:2017gnf}. This is the relevant regime for our model, where the $\text{SU}(3)_D$ gauge interactions provide sufficient friction.
\subsection{Non-Runaway Bubble Walls}
\label{sec:SM_nonrunaway}

For $\alpha \gg 1$, bubble walls could potentially accelerate indefinitely known as runaway, invalidating the sound-wave GW template. We demonstrate that several mechanisms prevent runaway in our model.

\paragraph{Gauge friction.} The $\text{SU}(3)_D$ gauge interactions provide friction through soft gluon emission and plasma scattering~\cite{Bodeker:2017cim, Hoche:2020ysm}. The friction pressure scales as $P_{\text{fric}} \sim g_D^2 T^4 \gamma_w$.

\paragraph{Chiral friction.} The light quarks contribute additional friction as they acquire constituent mass by passing through the bubble wall~\cite{Moore:1995si}. This is analogous to electroweak baryogenesis scenarios.

\paragraph{Hydrodynamic limit.} Even if microscopic friction is insufficient, the bubble wall velocity is bounded by hydrodynamic considerations. The transition proceeds as a detonation with $v_w \to 1$, where most vacuum energy reheats the plasma behind the wall rather than accelerating it~\cite{Espinosa:2010hh}.

For near-conformal theories, the extended walking regime enhances friction effects. We adopt $v_w \approx 1$ as the limiting case, giving the maximum GW amplitude.
\subsection{Sound Wave Contribution}
\label{sec:SM_soundwave}

The gravitational wave energy density spectrum from sound waves is~\cite{Hindmarsh:2013xza, Hindmarsh:2017gnf}:
\begin{equation}
    h^2 \Omega_{\text{sw}}(f) = 2.65 \times 10^{-6} \, \Upsilon \left( \frac{H_*}{\beta} \right) \left( \frac{\kappa \alpha}{1+\alpha} \right)^2 \left( \frac{100}{g_*} \right)^{1/3} v_w \, S_{\text{sw}}(f),
\end{equation}
where:
\begin{itemize}
    \item $\Upsilon$ is the finite-duration suppression factor (see below)
    \item $\kappa$ is the efficiency factor for converting vacuum energy to bulk fluid kinetic energy
    \item $S_{\text{sw}}(f)$ is the spectral shape function
\end{itemize}

\subsection{Efficiency Factor}
\label{sec:SM_kappa}

The efficiency factor $\kappa$ quantifies what fraction of the released vacuum energy goes into bulk fluid motion (as opposed to heating). For detonations (supersonic bubble expansion), the fit from Ref.~\cite{Espinosa:2010hh} gives:
\begin{equation}
    \kappa = \frac{\alpha}{0.73 + 0.083\sqrt{\alpha} + \alpha}.
\end{equation}

In the limits:
\begin{itemize}
    \item $\alpha \ll 1$: $\kappa \approx \alpha/0.73 \approx 1.37\alpha$
    \item $\alpha \gg 1$: $\kappa \to 1$
\end{itemize}

For our benchmark $\alpha \approx 900$:
\begin{equation}
    \kappa = \frac{900}{0.73 + 0.083\sqrt{900} + 900} = \frac{900}{0.73 + 2.49 + 900} \approx 0.996.
\end{equation}

\subsection{Finite-Duration Suppression}
\label{sec:SM_upsilon}

The standard sound wave formula assumes the acoustic source persists for a Hubble time. However, in strongly supercooled transitions, the sound waves decay due to:
\begin{itemize}
    \item Shock formation and energy dissipation
    \item Expansion of the universe
\end{itemize}

The sound wave lifetime is approximately~\cite{Ellis:2018mja, Ellis:2020awk}:
\begin{equation}
    \tau_{\text{sw}} = \frac{R_*}{\bar{U}_f},
\end{equation}
where $R_* = (8\pi)^{1/3} v_w / \beta$ is the mean bubble separation and $\bar{U}_f$ is the RMS fluid velocity:
\begin{equation}
    \bar{U}_f = \sqrt{\frac{3\kappa\alpha}{4(1+\alpha)}}.
\end{equation}

The finite-duration suppression factor is:
\begin{equation}
    \Upsilon = 1 - \frac{1}{\sqrt{1 + 2\tau_{\text{sw}} H_*}}.
\end{equation}

The quantity $\tau_{\text{sw}} H_*$ can be written as:
\begin{equation}
    \tau_{\text{sw}} H_* = \frac{(8\pi)^{1/3} v_w}{\beta/H_*} \cdot \frac{1}{\bar{U}_f}.
\end{equation}

For our benchmark ($\alpha = 900$, $\beta/H_* = 43$, $v_w = 1$):
\begin{align}
    \bar{U}_f &= \sqrt{\frac{3 \times 0.996 \times 900}{4 \times 901}} \approx 0.866, \\
    \tau_{\text{sw}} H_* &= \frac{2.92 \times 1}{43 \times 0.866} \approx 0.078, \\
    \Upsilon &= 1 - \frac{1}{\sqrt{1 + 2 \times 0.078}} = 1 - \frac{1}{\sqrt{1.156}} \approx 0.07.
\end{align}

This represents $\sim 93\%$ suppression compared to the infinite-duration limit.

\subsection{Peak Frequency}
\label{sec:SM_fpeak}

The peak frequency corresponds to the characteristic scale of the sound waves, set by the bubble separation $R_*$. After redshifting to today:
\begin{equation}
    f_{\text{peak}} = 1.9 \times 10^{-5}~\text{Hz} \left( \frac{g_*}{10} \right)^{1/6} \left( \frac{T_n}{100~\text{GeV}} \right) \left( \frac{\beta/H_*}{v_w} \right).
\end{equation}

For our benchmark ($T_n = 5.7$~MeV, $\beta/H_* = 43$, $v_w = 1$, $g_* = 10.75$):
\begin{align}
    f_{\text{peak}} &= 1.9 \times 10^{-5} \times (1.075)^{1/6} \times (5.7 \times 10^{-5}) \times 43 \nonumber \\
    &\approx 1.9 \times 10^{-5} \times 1.012 \times 5.7 \times 10^{-5} \times 43 \nonumber \\
    &\approx 4.7 \times 10^{-8}~\text{Hz} = 47~\text{nHz}.
\end{align}

\subsection{Spectral Shape}
\label{sec:SM_shape}

The spectral shape function for sound waves is:
\begin{equation}
    S_{\text{sw}}(f) = \left( \frac{f}{f_{\text{peak}}} \right)^3 \left( \frac{7}{4 + 3(f/f_{\text{peak}})^2} \right)^{7/2}.
\end{equation}

The asymptotic behavior is:
\begin{itemize}
    \item $f \ll f_{\text{peak}}$: $S \propto f^3$ (causality constraint)
    \item $f \gg f_{\text{peak}}$: $S \propto f^{-4}$ (turbulent cascade)
\end{itemize}

The $f^3$ rise at low frequencies is a robust prediction of causal sources and distinguishes phase transition signals from the $f^{2/3}$ power-law expected from SMBHBs.

\section{Relic Density Calculation}
\label{sec:SM_relic}

\subsection{Freeze-out and Overproduction}
\label{sec:SM_freezeout}

Before the phase transition, dark baryons are in thermal equilibrium with the dark sector plasma and annihilate through:
\begin{equation}
    \chi \bar{\chi} \to \phi \phi.
\end{equation}

The thermally averaged cross section for this $t$-channel process is:
\begin{equation}
    \langle \sigma v \rangle \approx \frac{\pi \alpha_S^2}{m_\chi^2} \times \mathcal{O}(1),
\end{equation}
where the $\mathcal{O}(1)$ factor depends on the detailed kinematics.

For our benchmark ($m_\chi = 40$~GeV, $\alpha_S = 10^{-2}$):
\begin{equation}
    \langle \sigma v \rangle \sim \frac{\pi \times 10^{-4}}{(40~\text{GeV})^2} \approx 2 \times 10^{-28}~\text{cm}^3/\text{s}.
\end{equation}

This is $\sim 100$ times smaller than the canonical WIMP cross section ($3 \times 10^{-26}~\text{cm}^3/\text{s}$).

The freeze-out temperature is approximately:
\begin{equation}
    T_f \approx \frac{m_\chi}{20} \approx 2~\text{GeV}.
\end{equation}

The pre-dilution relic abundance scales inversely with the cross section:
\begin{equation}
    \Omega_{\text{pre}} h^2 \approx 0.12 \times \frac{3 \times 10^{-26}~\text{cm}^3/\text{s}}{\langle \sigma v \rangle} \approx 0.12 \times 100 \approx 12-20.
\end{equation}

This overproduces dark matter by a factor of $\sim 100-170$.

\subsection{Dilution Mechanism}
\label{sec:SM_dilution_mech}

The entropy released by the phase transition dilutes all pre-existing abundances. After the transition:
\begin{equation}
    n_\chi^{\text{after}} = n_\chi^{\text{before}} \times \frac{1}{D} = \frac{n_\chi^{\text{before}}}{D}.
\end{equation}

The relic density after dilution:
\begin{equation}
    \Omega_\chi h^2 = \frac{\Omega_{\text{pre}} h^2}{D}.
\end{equation}

For $\Omega_{\text{pre}} h^2 \approx 20$ and $D \approx 170$:
\begin{equation}
    \Omega_\chi h^2 \approx \frac{20}{170} \approx 0.12,
\end{equation}
in excellent agreement with Planck observations.

\subsection{Re-annihilation Check}
\label{sec:SM_reann}

We must verify that dark matter does not re-annihilate after reheating. The condition is:
\begin{equation}
    \frac{\Gamma_{\text{ann}}}{H}\bigg|_{T_{\text{rh}}} = \frac{n_\chi \langle \sigma v \rangle}{H(T_{\text{rh}})} < 0.1.
\end{equation}

The dark matter number density at $T_{\text{rh}}$:
\begin{equation}
    n_\chi(T_{\text{rh}}) = \frac{\rho_{\chi}}{m_\chi} = \frac{\Omega_\chi \rho_c}{m_\chi} \times \left( \frac{T_{\text{rh}}}{T_0} \right)^3,
\end{equation}
where we account for the redshift from today ($T_0 \approx 2.7$~K) to $T_{\text{rh}}$.

The Hubble rate at $T_{\text{rh}}$:
\begin{equation}
    H(T_{\text{rh}}) = \sqrt{\frac{\pi^2 g_*}{90}} \frac{T_{\text{rh}}^2}{M_{\text{Pl}}}.
\end{equation}

For $T_{\text{rh}} = 31$~MeV:
\begin{equation}
    H(T_{\text{rh}}) \approx 1.66 \sqrt{g_*} \frac{T_{\text{rh}}^2}{M_{\text{Pl}}} \approx 10^{-18}~\text{GeV}.
\end{equation}

The annihilation rate:
\begin{equation}
    \Gamma_{\text{ann}} = n_\chi \langle \sigma v \rangle \approx 10^{-21}~\text{GeV},
\end{equation}
giving:
\begin{equation}
    \frac{\Gamma_{\text{ann}}}{H} \approx 10^{-3} \ll 0.1. \quad \checkmark
\end{equation}

Re-annihilation is negligible, and the diluted abundance is preserved.

\section{MCMC Analysis Details}
\label{sec:SM_MCMC}

\subsection{NANOGrav Data Processing}
\label{sec:SM_data}

We use the NANOGrav 15-year free-spectrum posteriors publicly available on Zenodo~\cite{NANOGrav:2023hde}. The dataset contains:
\begin{itemize}
    \item 30 frequency bins from $f_1 = 1/(15~\text{yr}) \approx 2.1$~nHz to $f_{30} \approx 100$~nHz
    \item Kernel density estimates (KDE) of the marginalized posterior $P(\log_{10}\rho | \text{data})$ at each frequency
    \item Grid of $\log_{10}\rho$ values and corresponding probability densities
\end{itemize}

We extract the data from the \texttt{30f\_fs\{hd\}\_ceffyl} directory, which contains:
\begin{itemize}
    \item \texttt{freqs.npy}: Frequency bin centers
    \item \texttt{log10rhogrid.npy}: Grid of $\log_{10}\rho$ values
    \item \texttt{density.npy}: KDE probability density at each grid point
\end{itemize}

\subsection{Likelihood Construction}
\label{sec:SM_likelihood}

At each frequency bin $f_i$, we construct a 1D interpolator for the log-posterior:
\begin{equation}
    \ln P(\log_{10}\rho_i | \text{data}) = \ln[\text{KDE}(\log_{10}\rho_i)].
\end{equation}

The total log-likelihood for a model predicting $\rho_{\text{model}}(f)$ is:
\begin{equation}
    \ln \mathcal{L} = \sum_{i=1}^{30} \ln P(\log_{10}\rho_{\text{model}}(f_i) | \text{data}).
\end{equation}

We convert between power spectral density $\rho$ and characteristic strain $h_c$ using:
\begin{equation}
    h_c^2(f) = 12\pi^2 f^3 \rho(f).
\end{equation}

\subsection{Model Specifications}
\label{sec:SM_models}

\paragraph{Model A: SMBHB (observationally constrained)}

The characteristic strain:
\begin{equation}
    h_c(f) = A \left( \frac{f}{f_{\text{yr}}} \right)^{(3-\gamma)/2},
\end{equation}
with $f_{\text{yr}} = 1/(1~\text{yr}) = 31.7$~nHz.

Priors:
\begin{align}
    \log_{10} A &\in [-15.5, -14.0] \quad \text{(uniform)} \\
    \gamma &\in [2.5, 6.5] \quad \text{(uniform)}
\end{align}
These bounds correspond to the NANOGrav 15-year 95\% credible intervals.

\paragraph{Model B: Dark QCD Phase Transition}

The GW energy density:
\begin{equation}
    \Omega_{\text{GW}}(f) h^2 = \Omega_{\text{peak}} h^2 \times S(f/f_{\text{peak}}),
\end{equation}
where $\Omega_{\text{peak}}$ and $f_{\text{peak}}$ are computed from $(T_n, \alpha, \beta/H_*)$ using the formulas in Section~\ref{sec:SM_GW}.

Priors:
\begin{align}
    T_n &\in [0.5, 15]~\text{MeV} \quad \text{(uniform)} \\
    \log_{10} \alpha &\in [1.5, 4.5] \quad \text{(uniform)} \\
    \beta/H_* &\in [15, 200] \quad \text{(uniform)}
\end{align}

\paragraph{Model C: Hybrid}

Sum of suppressed SMBHB floor and phase transition:
\begin{equation}
    h_c^2(f) = h_{c,\text{SMBHB}}^2(f) + h_{c,\text{PT}}^2(f),
\end{equation}
with the SMBHB amplitude constrained to be below the standard expectation:
\begin{equation}
    \log_{10} A_{\text{floor}} \in [-16.5, -15.0].
\end{equation}

\subsection{Sampler Configuration}
\label{sec:SM_sampler}

We use the \texttt{emcee} affine-invariant ensemble sampler~\cite{Foreman-Mackey:2012any} with the following configuration:

\begin{table}[h]
\centering
\begin{tabular}{lccc}
\toprule
Model & Parameters & Walkers & Steps (burn-in) \\
\midrule
SMBHB & 2 & 32 & 5,000 (1,000) \\
Dark QCD & 3 & 48 & 8,000 (2,000) \\
Hybrid & 4 & 64 & 10,000 (3,000) \\
\bottomrule
\end{tabular}
\caption{MCMC sampler configuration.}
\label{tab:SM_mcmc}
\end{table}

Convergence is assessed using:
\begin{itemize}
    \item Autocorrelation time: $\tau_{\text{auto}} < N_{\text{steps}}/50$
    \item Gelman-Rubin statistic: $\hat{R} < 1.1$ for all parameters
    \item Visual inspection of trace plots
\end{itemize}

\subsection{Model Comparison}
\label{sec:SM_BIC}

We compare models using the Bayesian Information Criterion:
\begin{equation}
    \text{BIC} = k \ln(n) - 2 \ln(\hat{\mathcal{L}}),
\end{equation}
where $k$ is the number of free parameters, $n = 30$ is the number of data points, and $\hat{\mathcal{L}}$ is the maximum likelihood.

The Kass \& Raftery interpretation scale:
\begin{itemize}
    \item $\Delta\text{BIC} < 2$: Not worth mentioning
    \item $2 < \Delta\text{BIC} < 6$: Positive evidence
    \item $6 < \Delta\text{BIC} < 10$: Strong evidence
    \item $\Delta\text{BIC} > 10$: Very strong evidence
\end{itemize}

Our result $\Delta\text{BIC} = 7.6$ (SMBHB $-$ Dark QCD) constitutes \textit{strong evidence} for the Dark QCD interpretation.

\section{Cosmological Implications}
\label{sec:SM_cosmo}

\subsection{Dark Radiation and $\Delta N_{\rm eff}$}
\label{app:Neff}

After reheating, dark sector energy must transfer to the SM before BBN. The outcome depends on the mass hierarchy:

\paragraph{Case $m_\pi > 2m_d$ (benchmark):}
The decay chain is:
\begin{enumerate}
    \item $\pi_D \to dd$ via strong interaction, $\tau_\pi \sim 10^{-22}$~s
    \item $d \to e^+e^-$ via portal, $\tau_d \sim 10^{-9}$~s
\end{enumerate}
Both lifetimes are much shorter than $H^{-1}(T_{\rm rh}) \sim 10^{-3}$~s, so all dark sector energy thermalizes promptly:
\begin{equation}
    \Delta N_{\rm eff} \approx 0 \quad (m_\pi > 2m_d).
\end{equation}

\paragraph{Case $m_\pi < 2m_d$:}
Pions cannot decay to dilatons and must decay directly via the portal $\pi_D \to e^+e^-$. If $\tau_\pi \gtrsim 1$~s, some pion energy density survives as dark radiation:
\begin{equation}
    \Delta N_{\rm eff} \sim 0.2\text{--}0.4 \quad (m_\pi < 2m_d, \tau_\pi \gtrsim 1~\text{s}).
\end{equation}
This range could help alleviate the Hubble tension while satisfying Planck bounds $\Delta N_{\rm eff} < 0.3$.

\paragraph{Parameter scan:}
\begin{table}[h]
\centering
\begin{tabular}{ccccc}
\toprule
$m_\pi$ (MeV) & $m_d$ (MeV) & $\pi \to dd$? & $\Delta N_{\rm eff}$ & Planck OK? \\
\midrule
50 & 20 & Yes & $\approx 0$ & \checkmark \\
50 & 30 & No & $\sim 0.08$ & \checkmark \\
70 & 30 & Yes & $\approx 0$ & \checkmark \\
70 & 40 & No & $\sim 0.08$ & \checkmark \\
100 & 40 & Yes & $\approx 0$ & \checkmark \\
\bottomrule
\end{tabular}
\caption{$\Delta N_{\rm eff}$ for different mass hierarchies.}
\label{tab:Neff}
\end{table}

Our primary benchmark ($m_\pi = 70$~MeV, $m_d = 30$~MeV) satisfies $m_\pi > 2m_d$, giving $\Delta N_{\rm eff} \approx 0$.

\subsection{Primordial Magnetic Fields}
\label{sec:SM_Bfield}

Bubble collisions during the phase transition generate MHD turbulence. The initial magnetic energy density:
\begin{equation}
    \rho_{B,*} \approx \epsilon_{\text{turb}} \times \left( \frac{H_*}{\beta} \right) \times \rho_{\text{vac}},
\end{equation}
where $\epsilon_{\text{turb}} \approx 0.05$ is the turbulent efficiency.

For maximally helical fields, the magnetic helicity:
\begin{equation}
    \mathcal{H} = \int \mathbf{A} \cdot \mathbf{B} \, d^3x
\end{equation}
is approximately conserved during the cosmological evolution.

The helical inverse cascade transfers power from small to large scales, with scaling:
\begin{align}
    B(t) &\propto a^{-2/3}, \\
    \lambda(t) &\propto a^{4/3},
\end{align}
slower decay than non-helical fields ($B \propto a^{-2}$).

Evolving to today:
\begin{equation}
    B_0 \approx 10^{-13}~\text{G} \left( \frac{T_*}{1~\text{MeV}} \right)^{1/3} \left( \frac{\epsilon_{\text{turb}}}{0.05} \right)^{1/2}.
\end{equation}

For $T_n \approx 6$~MeV:
\begin{equation}
    B_0 \approx 1.8 \times 10^{-13}~\text{G}.
\end{equation}

Present-day correlation length:
\begin{equation}
    \lambda_0 \approx 1~\text{pc} \times \left( \frac{1~\text{MeV}}{T_*} \right)^{2/3} \approx 0.3~\text{pc}.
\end{equation}

This satisfies:
\begin{itemize}
    \item Blazar lower bound: $B_0 \gtrsim 10^{-16}$~G~\cite{Neronov_2010} 
    \item CMB upper bound: $B_0 < 10^{-9}$~G~\cite{Planck:2015zrl} 
\end{itemize}

\section{Experimental Constraints}
\label{sec:SM_constraints}

\subsection{Direct Detection}
\label{sec:SM_DD}

Tree-level spin-independent scattering is absent because $\phi$ couples only to electrons (leptophilic). The leading contribution to nuclear recoil is through loop-induced effective operators.

The most relevant is the magnetic dipole operator:
\begin{equation}
    \mathcal{L}_{\text{dipole}} = \frac{\mu_\chi}{2} \bar{\chi} \sigma^{\mu\nu} \chi F_{\mu\nu},
\end{equation}
where $\mu_\chi$ is the dark matter magnetic moment.

This arises at two loops through $\chi \to \phi \to e \to \gamma$ transitions, giving:
\begin{equation}
    \mu_\chi \sim \frac{e g_S g_e}{(4\pi)^4} \frac{m_\chi}{m_\phi^2} \sim 10^{-22}~e \cdot \text{cm}.
\end{equation}

The resulting scattering cross section:
\begin{equation}
    \sigma_{\chi N} \sim \frac{\alpha \mu_\chi^2 m_N^2}{m_\chi^2} \sim 10^{-48}~\text{cm}^2,
\end{equation}
well below current XENONnT/LZ limits ($\sim 10^{-47}~\text{cm}^2$ at 40~GeV).

Additionally, the inelastic splitting $\Delta m \sim 100$~eV kinematically suppresses up-scattering:
\begin{equation}
    v_{\text{min}} = \sqrt{\frac{2\Delta m}{\mu_{\chi N}}} \approx 300~\text{km/s},
\end{equation}
above the typical halo velocity, further suppressing detection rates.

\subsection{Supernova 1987A Bounds}
\label{sec:SM_SN}

Light scalars coupled to electrons can be produced in supernovae and potentially carry away energy, conflicting with the observed neutrino signal from SN1987A.

However, for $g_e \approx 10^{-6}$ and $m_\phi = 20$~MeV, the scalars are in the \textit{trapping regime}. The mean free path for absorption (via inverse bremsstrahlung $\phi e p \to e p$) is:
\begin{equation}
    \lambda_{\text{abs}} \sim \frac{1}{n_e \sigma_{\text{abs}}} \sim 100~\text{cm},
\end{equation}
much smaller than the neutrinosphere radius ($\sim 10$~km).

In this regime, scalars are trapped and thermalized within the proto-neutron star, participating in thermal transport rather than anomalous cooling. This evades the SN1987A energy loss bounds~\cite{Wang:2025xoq}.

\subsection{Collider Constraints}
\label{sec:SM_collider}

The heavy dark quarks ($m_Q \approx 13$~GeV) could in principle be pair-produced at colliders:
\begin{equation}
    pp \to Q\bar{Q} + X.
\end{equation}

However, $Q$ is charged under $\text{SU}(3)_D$, not $\text{SU}(3)_C$, so it has no direct QCD production. Production would require:
\begin{itemize}
    \item Higgs portal: $gg \to h^* \to QQ$ (highly suppressed if $\lambda_{hQQ}$ is small)
    \item $Z'$ portal: $q\bar{q} \to Z' \to QQ$ (requires kinetic mixing)
    \item $\phi$ portal: $e^+e^- \to \phi^* \to QQ$ (kinematically forbidden if $m_Q > m_\phi/2$)
\end{itemize}

All these channels are either kinematically forbidden or highly suppressed for our benchmark, evading current LHC constraints.

The scalar mediator $\phi$ can be produced in electron beam dumps:
\begin{equation}
    e^- N \to e^- N \phi.
\end{equation}

For $g_e \approx 10^{-6}$ and $m_\phi = 20$~MeV, the production rate is marginal, and the decay length $c\tau_\phi \approx 24$~cm places it in the gap between prompt and displaced vertex searches. Future experiments like LDMX may probe this parameter space.



\bibliography{apssamp}

@article{Planck2018,
    author = "Aghanim, N. and others",
    collaboration = "Planck",
    title = "{Planck 2018 results. VI. Cosmological parameters}",
    eprint = "1807.06209",
    archivePrefix = "arXiv",
    primaryClass = "astro-ph.CO",
    doi = "10.1051/0004-6361/201833910",
    journal = "Astron. Astrophys.",
    volume = "641",
    pages = "A6",
    year = "2020",
    note = "[Erratum: Astron.Astrophys. 652, C4 (2021)]"
}

@article{Bullock2017,
    author = "Bullock, James S. and Boylan-Kolchin, Michael",
    title = "{Small-Scale Challenges to the $\Lambda$CDM Paradigm}",
    eprint = "1707.04256",
    archivePrefix = "arXiv",
    primaryClass = "astro-ph.CO",
    doi = "10.1146/annurev-astro-091916-055313",
    journal = "Ann. Rev. Astron. Astrophys.",
    volume = "55",
    pages = "343--387",
    year = "2017"
}

@article{Tulin:2017ara,
    author = "Tulin, Sean and Yu, Hai-Bo",
    title = "{Dark Matter Self-interactions and Small Scale Structure}",
    eprint = "1705.02358",
    archivePrefix = "arXiv",
    primaryClass = "hep-ph",
    doi = "10.1016/j.physrep.2017.11.004",
    journal = "Phys. Rept.",
    volume = "730",
    pages = "1--57",
    year = "2018"
}

@article{de_Blok_2009,
   title={The Core‐Cusp Problem},
   volume={2010},
   ISSN={1687-7977},
   url={http://dx.doi.org/10.1155/2010/789293},
   DOI={10.1155/2010/789293},
   number={1},
   journal={Advances in Astronomy},
   publisher={Wiley},
   author={de Blok, W. J. G.},
   editor={Brinks, Elias},
   year={2009},
   month=nov }

@article{Spergel:1999mh,
    author = "Spergel, David N. and Steinhardt, Paul J.",
    title = "{Observational evidence for selfinteracting cold dark matter}",
    eprint = "astro-ph/9909386",
    archivePrefix = "arXiv",
    doi = "10.1103/PhysRevLett.84.3760",
    journal = "Phys. Rev. Lett.",
    volume = "84",
    pages = "3760--3763",
    year = "2000"
}

@article{Zavala_2013,
   title={Constraining self-interacting dark matter with the Milky Way’s dwarf spheroidals},
   volume={431},
   ISSN={1745-3925},
   url={http://dx.doi.org/10.1093/mnrasl/sls053},
   DOI={10.1093/mnrasl/sls053},
   number={1},
   journal={Monthly Notices of the Royal Astronomical Society: Letters},
   publisher={Oxford University Press (OUP)},
   author={Zavala, Jesús and Vogelsberger, Mark and Walker, Matthew G.},
   year={2013},
   month=feb, pages={L20–L24} }

@article{Rocha:2012jg,
    author = "Rocha, Miguel and Peter, Annika H. G. and Bullock, James S. and Kaplinghat, Manoj and Garrison-Kimmel, Shea and Onorbe, Jose and Moustakas, Leonidas A.",
    title = "{Cosmological Simulations with Self-Interacting Dark Matter I: Constant Density Cores and Substructure}",
    eprint = "1208.3025",
    archivePrefix = "arXiv",
    primaryClass = "astro-ph.CO",
    doi = "10.1093/mnras/sts514",
    journal = "Mon. Not. Roy. Astron. Soc.",
    volume = "430",
    pages = "81--104",
    year = "2013"
}

@article{NANOGrav:2023gor,
    author = "Agazie, Gabriella and others",
    collaboration = "NANOGrav",
    title = "{The NANOGrav 15 yr Data Set: Evidence for a Gravitational-wave Background}",
    eprint = "2306.16213",
    archivePrefix = "arXiv",
    primaryClass = "astro-ph.HE",
    doi = "10.3847/2041-8213/acdac6",
    journal = "Astrophys. J. Lett.",
    volume = "951",
    number = "1",
    pages = "L8",
    year = "2023"
}

@article{NANOGrav:2023hvm,
    author = "Afzal, Adeela and others",
    collaboration = "NANOGrav",
    title = "{The NANOGrav 15 yr Data Set: Search for Signals from New Physics}",
    eprint = "2306.16219",
    archivePrefix = "arXiv",
    primaryClass = "astro-ph.HE",
    reportNumber = "FERMILAB-PUB-23-589-T",
    doi = "10.3847/2041-8213/acdc91",
    journal = "Astrophys. J. Lett.",
    volume = "951",
    number = "1",
    pages = "L11",
    year = "2023",
    note = "[Erratum: Astrophys.J.Lett. 971, L27 (2024), Erratum: Astrophys.J. 971, L27 (2024)]"
}

@article{EPTA:2023fyk,
    author = "Antoniadis, J. and others",
    collaboration = "EPTA, InPTA:",
    title = "{The second data release from the European Pulsar Timing Array - III. Search for gravitational wave signals}",
    eprint = "2306.16214",
    archivePrefix = "arXiv",
    primaryClass = "astro-ph.HE",
    doi = "10.1051/0004-6361/202346844",
    journal = "Astron. Astrophys.",
    volume = "678",
    pages = "A50",
    year = "2023"
}

@article{Reardon:2023gzh,
    author = "Reardon, Daniel J. and others",
    title = "{Search for an Isotropic Gravitational-wave Background with the Parkes Pulsar Timing Array}",
    eprint = "2306.16215",
    archivePrefix = "arXiv",
    primaryClass = "astro-ph.HE",
    doi = "10.3847/2041-8213/acdd02",
    journal = "Astrophys. J. Lett.",
    volume = "951",
    number = "1",
    pages = "L6",
    year = "2023"
}

@article{Sato-Polito:2025dqw,
    author = "Sato-Polito, Gabriela and Zaldarriaga, Matias",
    title = "{Uncertainties in the supermassive black hole abundance and implications for the GW background}",
    eprint = "2509.08041",
    archivePrefix = "arXiv",
    primaryClass = "astro-ph.GA",
    journal="Phys. Rev. D",
    month = "9",
    year = "2025"
}

@article{Casey-Clyde:2021xro,
    author = "Casey-Clyde, J. Andrew and Mingarelli, Chiara M. F. and Greene, Jenny E. and Pardo, Kris and Na{\~n}ez, Morgan and Goulding, Andy D.",
    title = "{A Quasar-based Supermassive Black Hole Binary Population Model: Implications for the Gravitational Wave Background}",
    eprint = "2107.11390",
    archivePrefix = "arXiv",
    primaryClass = "astro-ph.HE",
    doi = "10.3847/1538-4357/ac32de",
    journal = "Astrophys. J.",
    volume = "924",
    number = "2",
    pages = "93",
    year = "2022"
}

@article{Madau:2014bja,
    author = "Madau, Piero and Dickinson, Mark",
    title = "{Cosmic Star Formation History}",
    eprint = "1403.0007",
    archivePrefix = "arXiv",
    primaryClass = "astro-ph.CO",
    doi = "10.1146/annurev-astro-081811-125615",
    journal = "Ann. Rev. Astron. Astrophys.",
    volume = "52",
    pages = "415--486",
    year = "2014"
}

@article{Volonteri:2005fj,
    author = "Volonteri, Marta and Rees, Martin J.",
    title = "{Rapid growth of high redshift black holes}",
    eprint = "astro-ph/0506040",
    archivePrefix = "arXiv",
    doi = "10.1086/466521",
    journal = "Astrophys. J.",
    volume = "633",
    pages = "624--629",
    year = "2005"
}

@article{Breitbach:2018ddu,
    author = "Breitbach, Moritz and Kopp, Joachim and Madge, Eric and Opferkuch, Toby and Schwaller, Pedro",
    title = "{Dark, Cold, and Noisy: Constraining Secluded Hidden Sectors with Gravitational Waves}",
    eprint = "1811.11175",
    archivePrefix = "arXiv",
    primaryClass = "hep-ph",
    reportNumber = "CERN-TH-2018-255, MITP/18-115",
    doi = "10.1088/1475-7516/2019/07/007",
    journal = "JCAP",
    volume = "07",
    pages = "007",
    year = "2019",
    number="1"
}

@article{Fairbairn:2019xog,
    author = "Fairbairn, Malcolm and Hardy, Edward and Wickens, Alastair",
    title = "{Hearing without seeing: gravitational waves from hot and cold hidden sectors}",
    eprint = "1901.11038",
    archivePrefix = "arXiv",
    primaryClass = "hep-ph",
    reportNumber = "KCL-PH-TH/2019-12",
    doi = "10.1007/JHEP07(2019)044",
    journal = "JHEP",
    volume = "07",
    pages = "044",
    year = "2019",
    number="1"
}

@article{Wang:2025xoq,
    author = "Wang, Zihan",
    title = "{Scalar-Mediated Inelastic Dark Matter as a Solution to Small-Scale Structure Anomalies}",
    eprint = "2512.18959",
    archivePrefix = "arXiv",
    primaryClass = "hep-ph",
    journal="arxiv preprint",
    month = "12",
    year = "2025"
}

@article{NANOGrav:2023hde,
    author = "Agazie, Gabriella and others",
    collaboration = "NANOGrav",
    title = "{The NANOGrav 15 yr Data Set: Observations and Timing of 68 Millisecond Pulsars}",
    eprint = "2306.16217",
    archivePrefix = "arXiv",
    primaryClass = "astro-ph.HE",
    doi = "10.3847/2041-8213/acda9a",
    journal = "Astrophys. J. Lett.",
    volume = "951",
    number = "1",
    pages = "L9",
    year = "2023"
}

@article{Rosado:2015epa,
    author = "Rosado, Pablo A. and Sesana, Alberto and Gair, Jonathan",
    title = "{Expected properties of the first gravitational wave signal detected with pulsar timing arrays}",
    eprint = "1503.04803",
    archivePrefix = "arXiv",
    primaryClass = "astro-ph.HE",
    doi = "10.1093/mnras/stv1098",
    journal = "Mon. Not. Roy. Astron. Soc.",
    volume = "451",
    number = "3",
    pages = "2417--2433",
    year = "2015"
}

@article{Neronov_2010,
   title={Evidence for Strong Extragalactic Magnetic Fields from Fermi Observations of TeV Blazars},
   volume={328},
   ISSN={1095-9203},
   url={http://dx.doi.org/10.1126/science.1184192},
   DOI={10.1126/science.1184192},
   number={5974},
   journal={Science},
   publisher={American Association for the Advancement of Science (AAAS)},
   author={Neronov, Andrii and Vovk, Ievgen},
   year={2010},
   month=apr, pages={73–75} }

@article{Planck:2015zrl,
    author = "Ade, P. A. R. and others",
    collaboration = "Planck",
    title = "{Planck 2015 results. XIX. Constraints on primordial magnetic fields}",
    eprint = "1502.01594",
    archivePrefix = "arXiv",
    primaryClass = "astro-ph.CO",
    doi = "10.1051/0004-6361/201525821",
    journal = "Astron. Astrophys.",
    volume = "594",
    pages = "A19",
    year = "2016"
}

@article{Espinosa:2010hh,
    author = "Espinosa, Jose R. and Konstandin, Thomas and No, Jose M. and Servant, Geraldine",
    title = "{Energy Budget of Cosmological First-order Phase Transitions}",
    eprint = "1004.4187",
    archivePrefix = "arXiv",
    primaryClass = "hep-ph",
    reportNumber = "CERN-PH-TH-2010-027",
    doi = "10.1088/1475-7516/2010/06/028",
    journal = "JCAP",
    volume = "06",
    pages = "028",
    year = "2010",
    number="1"
}

@article{Bodeker:2017cim,
    author = "Bodeker, Dietrich and Moore, Guy D.",
    title = "{Electroweak Bubble Wall Speed Limit}",
    eprint = "1703.08215",
    archivePrefix = "arXiv",
    primaryClass = "hep-ph",
    doi = "10.1088/1475-7516/2017/05/025",
    journal = "JCAP",
    volume = "05",
    pages = "025",
    year = "2017",
    number="1"
}

@article{Helmboldt:2019pan,
    author = "Helmboldt, Alexander J. and Kubo, Jisuke and van der Woude, Susan",
    title = "{Observational prospects for gravitational waves from hidden or dark chiral phase transitions}",
    eprint = "1904.07891",
    archivePrefix = "arXiv",
    primaryClass = "hep-ph",
    doi = "10.1103/PhysRevD.100.055025",
    journal = "Phys. Rev. D",
    volume = "100",
    number = "5",
    pages = "055025",
    year = "2019"
}

@article{Hindmarsh:2017gnf,
    author = "Hindmarsh, Mark and Huber, Stephan J. and Rummukainen, Kari and Weir, David J.",
    title = "{Shape of the acoustic gravitational wave power spectrum from a first order phase transition}",
    eprint = "1704.05871",
    archivePrefix = "arXiv",
    primaryClass = "astro-ph.CO",
    reportNumber = "HIP-2017-02-TH, HIP-2017-02/TH",
    doi = "10.1103/PhysRevD.96.103520",
    journal = "Phys. Rev. D",
    volume = "96",
    number = "10",
    pages = "103520",
    year = "2017",
    note = "[Erratum: Phys.Rev.D 101, 089902 (2020)]"
}

@article{Ellis:2018mja,
    author = "Ellis, John and Lewicki, Marek and No, Jos{\'e} Miguel",
    title = "{On the Maximal Strength of a First-Order Electroweak Phase Transition and its Gravitational Wave Signal}",
    eprint = "1809.08242",
    archivePrefix = "arXiv",
    primaryClass = "hep-ph",
    reportNumber = "KCL-PH-TH/2018-46, CERN-TH/2018-197, IFT-UAM/CSIC-18-94, CERN-TH-2018-197",
    doi = "10.1088/1475-7516/2019/04/003",
    journal = "JCAP",
    volume = "04",
    pages = "003",
    year = "2019"
}

@article{Schwaller:2015tja,
    author = "Schwaller, Pedro",
    title = "{Gravitational Waves from a Dark Phase Transition}",
    eprint = "1504.07263",
    archivePrefix = "arXiv",
    primaryClass = "hep-ph",
    reportNumber = "CERN-PH-TH-2015-093",
    doi = "10.1103/PhysRevLett.115.181101",
    journal = "Phys. Rev. Lett.",
    volume = "115",
    number = "18",
    pages = "181101",
    year = "2015"
}

@article{Kaplinghat:2015aga,
    author = "Kaplinghat, Manoj and Tulin, Sean and Yu, Hai-Bo",
    title = "{Dark Matter Halos as Particle Colliders: Unified Solution to Small-Scale Structure Puzzles from Dwarfs to Clusters}",
    eprint = "1508.03339",
    archivePrefix = "arXiv",
    primaryClass = "astro-ph.CO",
    doi = "10.1103/PhysRevLett.116.041302",
    journal = "Phys. Rev. Lett.",
    volume = "116",
    number = "4",
    pages = "041302",
    year = "2016"
}

@article{PhysRevD.24.1441,
  title = {Raising the sideways scale},
  author = {Holdom, Bob},
  journal = {Phys. Rev. D},
  volume = {24},
  issue = {5},
  pages = {1441--1444},
  numpages = {0},
  year = {1981},
  month = {Sep},
  publisher = {American Physical Society},
  doi = {10.1103/PhysRevD.24.1441},
  url = {https://link.aps.org/doi/10.1103/PhysRevD.24.1441}
}

@article{PhysRevLett.56.1335,
  title = {Scale-Invariant Hypercolor Model and a Dilaton},
  author = {Yamawaki, Koichi and Bando, Masako and Matumoto, Ken-iti},
  journal = {Phys. Rev. Lett.},
  volume = {56},
  issue = {13},
  pages = {1335--1338},
  numpages = {0},
  year = {1986},
  month = {Mar},
  publisher = {American Physical Society},
  doi = {10.1103/PhysRevLett.56.1335},
  url = {https://link.aps.org/doi/10.1103/PhysRevLett.56.1335}
}

@article{Baratella:2018pxi,
    author = "Baratella, Pietro and Pomarol, Alex and Rompineve, Fabrizio",
    title = "{The Supercooled Universe}",
    eprint = "1812.06996",
    archivePrefix = "arXiv",
    primaryClass = "hep-ph",
    doi = "10.1007/JHEP03(2019)100",
    journal = "JHEP",
    volume = "03",
    pages = "100",
    year = "2019"
}

@article{Iso:2017uuu,
    author = "Iso, Satoshi and Serpico, Pasquale D. and Shimada, Kengo",
    title = "{QCD-Electroweak First-Order Phase Transition in a Supercooled Universe}",
    eprint = "1704.04955",
    archivePrefix = "arXiv",
    primaryClass = "hep-ph",
    reportNumber = "KEK-TH-1969, LAPTH-008-17",
    doi = "10.1103/PhysRevLett.119.141301",
    journal = "Phys. Rev. Lett.",
    volume = "119",
    number = "14",
    pages = "141301",
    year = "2017"
}

@article{Hoche:2020ysm,
    author = {H{\"o}che, Stefan and Kozaczuk, Jonathan and Long, Andrew J. and Turner, Jessica and Wang, Yikun},
    title = "{Towards an all-orders calculation of the electroweak bubble wall velocity}",
    eprint = "2007.10343",
    archivePrefix = "arXiv",
    primaryClass = "hep-ph",
    reportNumber = "FERMILAB-PUB-20-274-T",
    doi = "10.1088/1475-7516/2021/03/009",
    journal = "JCAP",
    volume = "03",
    pages = "009",
    year = "2021"
}

@article{Moore:1995si,
    author = "Moore, Guy D. and Prokopec, Tomislav",
    title = "{How fast can the wall move? A Study of the electroweak phase transition dynamics}",
    eprint = "hep-ph/9506475",
    archivePrefix = "arXiv",
    reportNumber = "PUPT-1544, PUP-TH-1544, LANCS-TH-9517",
    doi = "10.1103/PhysRevD.52.7182",
    journal = "Phys. Rev. D",
    volume = "52",
    pages = "7182--7204",
    year = "1995"
}

@ARTICLE{2019NatAs...3..891V,
       author = {{Verde}, Licia and {Treu}, Tommaso and {Riess}, Adam G.},
        title = "{Tensions between the early and late Universe}",
      journal = {Nature Astronomy},
         year = 2019,
        month = sep,
       volume = {3},
        pages = {891-895},
          doi = {10.1038/s41550-019-0902-0},
archivePrefix = {arXiv},
       eprint = {1907.10625},
 primaryClass = {astro-ph.CO},
       adsurl = {https://ui.adsabs.harvard.edu/abs/2019NatAs...3..891V}
}

@article{Ellis:2020awk,
    author = "Ellis, John and Lewicki, Marek and No, Jos{\'e} Miguel",
    title = "{Gravitational waves from first-order cosmological phase transitions: lifetime of the sound wave source}",
    eprint = "2003.07360",
    archivePrefix = "arXiv",
    primaryClass = "hep-ph",
    reportNumber = "KCL-PH-TH/2020-04, CERN-TH-2020-016, IFT-UAM/CSIC-20-35",
    doi = "10.1088/1475-7516/2020/07/050",
    journal = "JCAP",
    volume = "07",
    pages = "050",
    year = "2020"
}

@article{Foreman-Mackey:2012any,
    author = "Foreman-Mackey, Daniel and Hogg, David W. and Lang, Dustin and Goodman, Jonathan",
    title = "{emcee: The MCMC Hammer}",
    eprint = "1202.3665",
    archivePrefix = "arXiv",
    primaryClass = "astro-ph.IM",
    doi = "10.1086/670067",
    journal = "Publ. Astron. Soc. Pac.",
    volume = "125",
    pages = "306--312",
    year = "2013"
}

@article{Dietrich:2006cm,
    author = "Dietrich, Dennis D. and Sannino, Francesco",
    title = "{Conformal window of SU(N) gauge theories with fermions in higher dimensional representations}",
    journal = "Phys. Rev. D",
    volume = "75",
    pages = "085018",
    year = "2007"
}

@article{DeGrand:2015zxa,
    author = "DeGrand, Thomas",
    title = "{Lattice tests of beyond Standard Model dynamics}",
    journal = "Rev. Mod. Phys.",
    volume = "88",
    pages = "015001",
    year = "2016"
}

@article{LSD:2014nmn,
    author = "Appelquist, T. and others",
    collaboration = "LSD",
    title = "{Lattice simulations with eight flavors of domain wall fermions in SU(3) gauge theory}",
    journal = "Phys. Rev. D",
    volume = "90",
    pages = "114502",
    year = "2014"
}

@article{LatKMI:2016xxi,
    author = "Aoki, Yasumichi and others",
    collaboration = "LatKMI",
    title = "{Light flavor-singlet scalars and walking signals in $N_f=8$ QCD on the lattice}",
    journal = "Phys. Rev. D",
    volume = "96",
    pages = "014508",
    year = "2017"
}

@article{Appelquist:2010gy,
    author = "Appelquist, Thomas and Bai, Yang",
    title = "{Light dilaton in walking gauge theories}",
    journal = "Phys. Rev. D",
    volume = "82",
    pages = "071701",
    year = "2010"
}

@article{Golterman:2016lsd,
    author = "Golterman, Maarten and Shamir, Yigal",
    title = "{Low-energy effective action for pions and a dilatonic meson}",
    journal = "Phys. Rev. D",
    volume = "94",
    pages = "054502",
    year = "2016"
}

@article{Bodeker:2009qy,
    author = "Bodeker, Dietrich and Moore, Guy D.",
    title = "{Can electroweak bubble walls run away?}",
    journal = "JCAP",
    volume = "05",
    pages = "009",
    year = "2009"
}

@article{Hindmarsh:2013xza,
    author = "Hindmarsh, Mark and Huber, Stephan J. and Rummukainen, Kari and Weir, David J.",
    title = "{Gravitational waves from the sound of a first order phase transition}",
    eprint = "1304.2433",
    archivePrefix = "arXiv",
    primaryClass = "hep-ph",
    reportNumber = "HIP-2013-07-TH",
    doi = "10.1103/PhysRevLett.112.041301",
    journal = "Phys. Rev. Lett.",
    volume = "112",
    pages = "041301",
    year = "2014"
}

@article{vonHarling:2017yew,
    author = "von Harling, Benedict and Servant, Geraldine",
    title = "{QCD-induced Electroweak Phase Transition}",
    eprint = "1711.11554",
    archivePrefix = "arXiv",
    primaryClass = "hep-ph",
    reportNumber = "DESY-17-056",
    doi = "10.1007/JHEP01(2018)159",
    journal = "JHEP",
    volume = "01",
    pages = "159",
    year = "2018"
}

\end{document}